\documentclass[12pt]{iopart}

\usepackage{iopams}  
\usepackage{graphicx}

\begin{document}
\title{Negation of photon loss provided by negative weak value}

\author{Kazuhiro Yokota  and Nobuyuki Imoto}
\address{Graduate School of Science, Osaka University,
Toyonaka, Osaka 560-0043, Japan
\footnote{This work was performed at Graduate School of Engineering Science, Osaka University.}
}

\ead{yokota@qi.mp.es.osaka-u.ac.jp}

\date{\today}

\begin{abstract}
We propose a usage of a weak value for a quantum processing between preselection and postselection.
While the weak value of a projector of 1 provides a process with certainty like the probability of $1$, the weak value of $-1$ negates the process completely.
Their mutually opposite effect is approved without a conventional `weak' condition.
In addition the quantum process is not limited to be unitary; in particular we consider a loss of photons and experimentally demonstrate the negation of the photon loss by using the negative weak value of $-1$ against the positive weak value of $1$.
\end{abstract}


\maketitle
\section{Introduction}
\label{sec:intro}
It is referred to as pre-postselection that both the initial and final states of a quantum system are selected on purpose.
Such a technique has been often made use of in quantum information to achieve state preparation, quantum processing, and so on.
The components which act on the quantum system are set up between the pre-postselection, and we obtain the quantum state we want after the pre-postselection.
We can determine how to design their arrangements and verify whether they accomplish the expected quantum processing by calculating the time evolution of the quantum system conventionally.

\begin{figure}
 \begin{center}
	\includegraphics[scale=0.9]{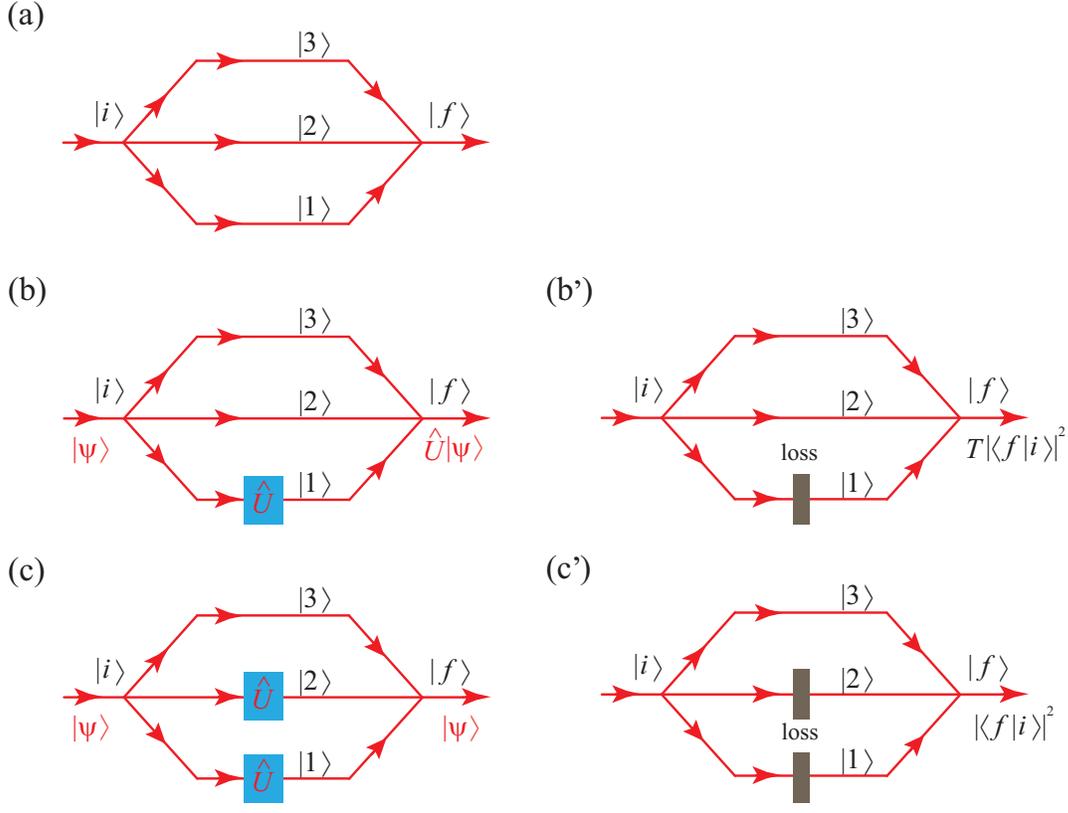}
 \end{center}
	\caption{(a)A photon can take the three paths $|1\rangle$, $|2\rangle$, and $|3\rangle$, the state of which is preselected in $|i\rangle$ and postselected in $|f\rangle$.
(b)When the unitary operator for the polarization, $\hat{U}$, is placed on the path of $|1\rangle$, the polarization initially in $|\psi\rangle$ is changed to $\hat{U}|\psi\rangle$ as if the photon takes the path with certainty.
(c)The same unitary operator is also set up on the path of $|2\rangle$. Then their operations cancel each other, that is, the polarization is $|\psi\rangle$ after the postselection.
(b')As in the case of (b), there is the device causing photon loss only on the path of $|1\rangle$, which has the transmission probability, $T$, for a photon.
Then, according to the transmission probability, the success probability of the postselection is given by $T|\langle f|i\rangle |^2$.
(c')In a similar manner, this loss of the success probability is negated by locating the same device on the path of $|2\rangle$ additionally.
}
\label{fig:intro}
\end{figure}

For example, we consider a photon whose path takes a superposition state in $|i\rangle =(|1\rangle +|2\rangle +|3\rangle )/\sqrt{3}$ initially as shown in figure \ref{fig:intro} (a).
Our goal is to operate $\hat{U}$ on its polarization of $|\psi\rangle$, which is placed only on the path of $|1\rangle$ in figure \ref{fig:intro} (b): we want the polarization of $\hat{U}|\psi\rangle$.
This can be easily performed, if we postselect the path state in $|f\rangle =(|1\rangle -|2\rangle +|3\rangle)/\sqrt{3}$.
In fact the time evolution of the photon is given as follows,
\begin{eqnarray}
|i\rangle |\psi\rangle &\rightarrow& (\hat{U}|1\rangle\langle 1|i\rangle +|2\rangle\langle 2|i\rangle +|3\rangle\langle 3|i\rangle)|\psi\rangle  \nonumber \\
&\rightarrow& (\hat{U}\langle f|1\rangle\langle 1|i\rangle-\langle f|2\rangle\langle 2|i\rangle+\langle f|3\rangle\langle 3|i\rangle)|\psi\rangle  \nonumber \\
&=& \langle f|i\rangle (\hat{U}-\hat{I}+\hat{I})|\psi\rangle = \langle f|i\rangle\hat{U}|\psi\rangle,
\end{eqnarray}
where $\hat{I}$ is an identity operator.
Under the condition of the success of postselection with the probability, $|\langle f|i\rangle|^2=1/9$, we can assure that the photon is certainly operated by $\hat{U}$ to be $\hat{U}|\psi\rangle$ as if it has passed $|1\rangle$ with certainty.
In other words, a completely destructive interference between the unexpected terms, namely, $-\hat{I}+\hat{I}$ is performed by the postselection at the cost of some samples which are failed in the postselection. 

On the contrary, we would like to consider how to cancel the operation of $\hat{U}$ in this pre-postselection.
Such a negation can be performed by having the same operation of $\hat{U}$ in $|2\rangle$ as shown in figure \ref{fig:intro} (c).
Actually, after the pre-postselection, the polarization of the photon is given as follows,
\begin{eqnarray}
|i\rangle |\psi\rangle &\rightarrow& (\hat{U}|1\rangle\langle 1|i\rangle +\hat{U}|2\rangle\langle 2|i\rangle +|3\rangle\langle 3|i\rangle)|\psi\rangle \nonumber \\
&\rightarrow& (\hat{U}\langle f|1\rangle\langle 1|i\rangle-\hat{U}\langle f|2\rangle\langle 2|i\rangle+\langle f|3\rangle\langle 3|i\rangle)|\psi\rangle \nonumber \\
&=& \langle f|i\rangle (\hat{U}-\hat{U}+\hat{I})|\psi\rangle = \langle f|i\rangle |\psi\rangle.
\end{eqnarray}

We have discussed how the polarization is operated by the pre-postselection of the path state, and the operation might be suggested as a unitary operation.
However, the similar situation will be conceivable, even if the operation is not unitary like photon loss.
The photon loss with the transmission probability, $T$, is put only on $|1\rangle$ in figure \ref{fig:intro} (b'), where we assume that this device does not introduce an additional phase to a photon.
We consider the success probability of the postselection, and how it is affected by the loss laying on the path.
Note that we do not care about the polarization in this case.
While, without the loss, the success probability of the postselection is given by $|\langle f|i\rangle |^2$, we can easily find that it drops to $T|\langle f|i\rangle |^2$ as follows,
\begin{eqnarray}
|i\rangle &\rightarrow& \sqrt{T}|1\rangle\langle 1|i\rangle +|2\rangle\langle 2|i\rangle +|3\rangle\langle 3|i\rangle \nonumber \\
&\rightarrow&  \sqrt{T}\langle f|1\rangle\langle 1|i\rangle +\langle f|2\rangle\langle 2|i\rangle +\langle f|3\rangle\langle 3|i\rangle \nonumber \\
&=& (\sqrt{T}-1+1)\langle f|i\rangle = \sqrt{T}\langle f|i\rangle.
\end{eqnarray}
That is to say, the success probability decreases as if photons certainly pass $|1\rangle$ and suffer the loss directly, which resembles the previous case in figure \ref{fig:intro} (b).

In like wise, the photon loss is negated, if we add the same device in $|2\rangle$ as shown in figure \ref{fig:intro} (c'), where we also assume that these two losses do not add some additional phases to a photon.
The success probability of the postselection gets back to $|\langle f|i\rangle |^2$ as follows,
\begin{eqnarray}
|i\rangle &\rightarrow& \sqrt{T}|1\rangle\langle 1|i\rangle +\sqrt{T}|2\rangle\langle 2|i\rangle +|3\rangle\langle 3|i\rangle \nonumber \\
&\rightarrow& \sqrt{T}\langle f|1\rangle\langle 1|i\rangle +\sqrt{T}\langle f|2\rangle\langle 2|i\rangle +\langle f|3\rangle\langle 3|i\rangle \nonumber \\
&=& (\sqrt{T}-\sqrt{T}+1)\langle f|i\rangle = \langle f|i\rangle. \label{eq:neg}
\end{eqnarray}

As we have seen, the above discussion on how the device put on each path participates in the state of a photon can be explained by calculating the time evolution conventionally.
On following the time evolution, it will be easy to be aware of that the effect of a device is given with a weight like $\langle f|1\rangle\langle 1|i\rangle$.
For example, in equation (\ref{eq:neg}), the negation of photon loss comes from the sum of $\langle f|1\rangle\langle 1|i\rangle$,  $\langle f|2\rangle\langle 2|i\rangle$, and $\langle f|3\rangle\langle 3|i\rangle$ normalized by $\langle f|i\rangle$ to be $\sqrt{T}-\sqrt{T}+1=1$.
Actually $\langle f|1\rangle\langle 1|i\rangle/\langle f|i\rangle$ is known as the weak value of $|1\rangle\langle 1|$, which characterizes a pre-postselected system \cite{W1, W2}; a weak value is uniquely determined by a preselection and a postselection.

In this paper, we show that a weak value gives us another manner to estimate the contributions of operations/losses between a pre-postselection, which is simpler than following a time evolution.
By associating a weak value with such a practical sense, we also clarify a meaning of a weak value itself; 
in particular a negative weak value represents a negation of operations/losses very well (Section \ref{subsec:theory}).
In fact we also experimentally demonstrate the negation of photon loss as mentioned in figure \ref{fig:intro} (c') (Section \ref{subsec:exp}).
Our discussion can be applied to a quantum system of not only one particle but also more than two particles, in which joint weak values appear;
we take up a specific case of two particles in Section \ref{sec:joint}.
Section \ref{sec:con} is devoted to our conclusion.

\section{Negation provided by negative weak value}
\subsection{Theory}
\label{subsec:theory}
First we would like to review weak value and how the value is useful in figure \ref{fig:intro}, which has been actually known as one of the application of weak value so far, namely, the quantum box problem \cite{QP1, QP2}.

Weak value was introduced as a result of weak measurement by Aharonov, Albert, and Vaidman \cite{W1, W2}.
While it has given us a new insight in foundation of quantum physics \cite{QF1}-\cite{QF9}, many applications of weak value has been also proposed \cite{W3}-\cite{W9}.
Given both an initial state, $|i\rangle$, and a final state, $|f\rangle$, the weak value of an observable, $\hat{O}$, is defined as follows,
\begin{eqnarray}
\langle\hat{O}\rangle_{\bf w} = \langle f|\hat{O}|i\rangle /\langle f|i\rangle .
\end{eqnarray}

In figure \ref{fig:intro} (a), the weak values of the projectors of $|1\rangle\langle 1|$, $|2\rangle\langle 2|$, and $|3\rangle\langle 3|$ can be calculated as follows,
\begin{eqnarray}
\langle |1\rangle\langle 1|\rangle_{\bf w}=\langle |3\rangle\langle 3|\rangle_{\bf w}=1, \ \langle |2\rangle\langle 2|\rangle_{\bf w}=-1.   \label{eq:wv_3box}
\end{eqnarray}
As the projector conventionally gives the probability for a photon to pass each path, it seems to be paradoxical: 
it is as if both of $|1\rangle$ and $|3\rangle$ have the probability 1.
In fact this case is known as the quantum box problem, which is one of the famous quantum paradoxes \cite{QP1, QP2}.
Then the other weak value, $\langle |2\rangle\langle 2|\rangle_{\bf w} =-1$, circumvents the paradox, as it satisfies the summation rule like conventional probabilities,
\begin{eqnarray}
\langle |1\rangle\langle 1|\rangle_{\bf w}+\langle |2\rangle\langle 2|\rangle_{\bf w}+\langle |3\rangle\langle 3|\rangle_{\bf w}=1.
\end{eqnarray}
Weak values have been applied to many quantum paradoxes so far \cite{QPj2}-\cite{QP9};
an appearance of the negative weak value $-1$ often plays an important role.

Although the weak value of $-1$ cannot be interpreted as a conventional probability, any weak value can be observed by weak measurements;
it also naturally appears in a quantum phenomenon as a value of a physical quantity \cite{W_ph1}-\cite{W_ph5}.
Recently we have also showed that positive and negative weak values are associated with symmetrical operations \cite{WV_sym}.

To return to our subject, it will be more clear what a weak value means practically, especially what a negative weak value represents.
As we discussed in the previous section, in figure \ref{fig:intro} (b) and (b'), the operation and the loss are performed as if a photon passes $|1\rangle$ with probability $1$.
That is to say, the weak value of $\langle |1\rangle\langle 1|\rangle_{\bf w}=1$ agrees with the probability of $1$.
In figure \ref{fig:intro} (c) and (c'), the same operation/loss is added on the path $|2\rangle$, and the operation/loss on $|1\rangle$ is negated.
Actually the sum of these weak values, $\langle|1\rangle\langle 1|\rangle_{\bf w}+\langle|2\rangle\langle 2|\rangle_{\bf w}=0$, corresponds to no operation/loss (i.e. the probability of $0$) (see also \cite{1-1}).
In this sense the negative weak value of $-1$ represents the negation very well, and plays a counterpart of the positive weak value of $1$, equivalently, the probability of $1$.
Especially, in the case of the loss with transmission probability $T=0$ (i.e. completely blocking by a shutter), blocking the path of the negative weak value of $-1$ corresponds to having a photon pass contrary.

\begin{figure}
 \begin{center}
	\includegraphics[scale=0.8]{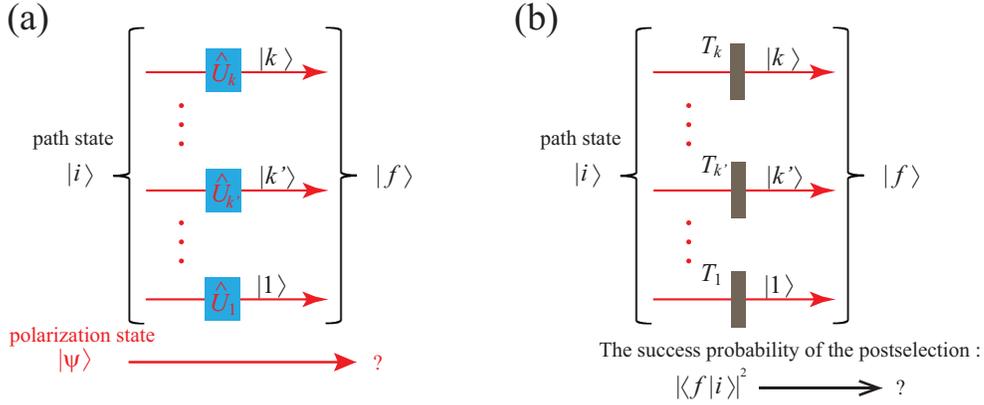}
 \end{center}
	\caption{A photon takes a superposition state of paths, which is pre-postselected.
(a)A component which operates the polarization of a photon is in each path.
(b)A loss with a transmission probability of a photon is in each path.
In both cases, the effect of the operation/loss is given with the weight of the weak value corresponding to each path:
the eventual polarization/the probability of the postselection is determined by the weak values.
}
\label{fig:intro2}
\end{figure}

In general the path state takes a superposition of $k$ paths, and the unitary operation of $\hat{U}_{k'}$ is placed on the $k'$th path as shown in figure \ref{fig:intro2} (a).
Then, after the pre-postselection on the path state, the polarization in initially $|\psi\rangle$ evolves as follows,
\begin{eqnarray}
|\psi\rangle \ \longrightarrow \ \frac{1}{\sqrt{N}}\sum_{k'=1}^{k} \langle |k'\rangle\langle k'|\rangle _{\bf w}\hat{U}_{k'}|\psi\rangle,    \label{eq:uni}
\end{eqnarray}
where $N$ represents the normalization of the polarization state, and the success probability of the postselection is given by $N|\langle f|i\rangle|^2$.
Note that the weak values should satisfy $\sum_{k'} \langle |k'\rangle\langle k'|\rangle _{\bf w}=1$ as conventional probabilities, because of $\sum_{k'} |k'\rangle\langle k'| =1$.

Correspondingly the loss with the transmission probability $T_{k'}$ is put in the $k'$th path in figure \ref{fig:intro2} (b).
In the pre-postselection on the path state, the success probability of the postselection is changed as follows,
\begin{eqnarray}
|\langle f|i\rangle|^2 \ \longrightarrow \ |\langle f|i\rangle|^2\left|\sum_{k'=1}^{k}\langle|k'\rangle\langle k'|\rangle_{\bf w}\sqrt{T_{k'}}\right|^2.   \label{eq:prob}
\end{eqnarray}

Clearly the negation takes place with a positive weak value $p(>0)$ and a negative weak value $-p$.
In other words the negation comes of the summation of these weak values, namely, $0$.
Actually, as the weak value $1$ has been regarded as the probability $1$, the weak value $0$ also corresponds to the probability $0$ (i.e. no operations/losses).
The weak values, $0$ and $1$, always satisfy such a definite relation with the probabilities \cite{QP1}, while it has been recently shown that the system with a weak value affects the other system as if it were in an eigenstate with eigenvalue equal to the weak value under the weak correlation \cite{WV3}.
In this sense the weak value of $-1$ has an important implication in relation to these definite weak values: 
$-1$ is the value which negates the definite value of $1$ (probability $1$) to be the definite value of $0$ (probability $0$).

In any case, without following the time evolution of a quantum state, we can derive the eventual outcome of the pre-postselection by using the weak value.
On purpose to perform our expected quantum process between pre-postselection, what we should do is to determine the weak values.
Although the weak values depend on both the initial and final states, each state itself does not matter.
There are many choices of these states to prepare the weak values we want.

Actually we experimentally demonstrated the negation provided by a negative weak value.
Before proceeding to the discussion, we would like to point out the difference from our previous result \cite{WV_sym} here.
First, a weak value appears without `weak' condition this time.
In our previous work, we also considered the situation like in figure \ref{fig:intro} (b); 
we discussed how the direction of linear-polarization of a photon is changed by the weak value of a projector on a path state.
As a result, we showed that the shift-angle given by a negative weak value of $-1$ is symmetrical to that one by a positive weak value of $1$, when the correlation between the path state and the polarization is weak.
This `weak' correlation follows the way `weak' measurement provides a weak value.
On the other hand, without such an approximation of weak correlation, we have shown that a weak value is associated with an actual operation this time.
Furthermore this result is not the generalization of our previous work (see Appendix).

Second, in the case of the loss on the path, the effect of a weak value directly appears in the probability of a pre-postselection.
In many research of weak value, a weak value has been verified in another degree of freedom like an operation of polarization in figure \ref{fig:intro} (b) and (c).
Actually, in weak measurement, a pointer shifts in response to a weak value given by the quantum system to be measured.
Our case indicates an application for calculating the behavior of a pre-postselected quantum system \cite{PP1}.

\subsection{Experiment - Method and Result}
\label{subsec:exp}
\begin{figure}
 \begin{center}
	\includegraphics[scale=0.7]{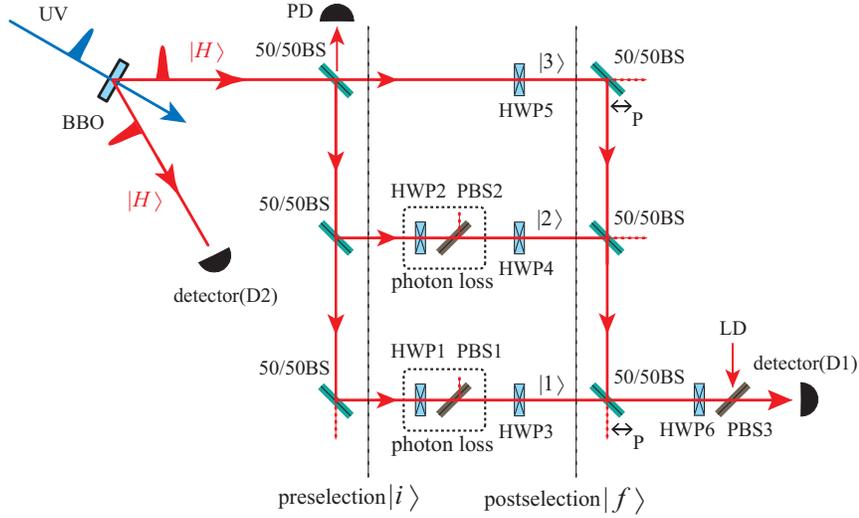}
 \end{center}
	\caption{Schematic of experimental implementation.
Horizontally polarized photon pairs were generated via spontaneous parametric down-conversion from type I phase matched BBO crystal pumped by a UV pulse (a central wavelength of 395nm and an average power of 120mW).
The UV pulse is taken from the frequency-doubled Ti:sapphire laser (wavelength of 790nm, pulse width of 100fs, and repetition rate of 80MHz).
While one of the photon pair was detected at D2 as just a trigger for a coincidence count, the other photon was incident to Mach-Zehnder interferometers composed of six 50/50 BSs.
Then the path state of the photon resulted in a superposition of $|1\rangle$, $|2\rangle$, and $|3\rangle$.
The laser diode (LD) and the photodiode (PD) were used to control the path lengths of Mach-Zehnder interferometer to achieve an appropriate postselection on the path state.
The dashed box composed of a half wave plate (HWP) and a polarized beam splitter (PBS) performed a loss of a photon, which gave a transmission probability we wanted.
The other HWPs were for weak measurement to observe the weak value of each path.
}
\label{fig:exp}
\end{figure}

We have experimentally performed the negation of the loss corresponding to figure \ref{fig:intro} (b') and (c').
Figure \ref{fig:exp} shows our experimental implementation composed of three Mach-Zehnder interferometers ($|1\rangle$ and $|2\rangle$, $|2\rangle$ and $|3\rangle$, $|3\rangle$ and $|1\rangle$), the visibilities of which were at least 98.3 $\pm$ 1.8\% for horizontally polarized photons.
The path lengths of the Mach-Zehnder interferometer composed of $|1\rangle$ and $|3\rangle$ ($|2\rangle$ and $|3\rangle$) were adjusted so that the port of D1 became the bright (dart) port, which was controlled by using the extra laser diode (LD) and the photodiode (PD) \cite{QPj4, WV_sym}:
in our experiment, we stopped counting photon at D1 every 5 seconds, and observed the fringe pattern of lights coming from LD at PD by moving the 50/50 BSs (beam splitters with reflectivity/transmissivity of 50\%) on the piezo stages (P).
Then the location of the BSs were reset to achieve the appropriate path lengths of interferometers as we have mentioned, and restarted the photon counting.
In this case, the path state of an incident photon initially in $|i\rangle=1/(2\sqrt{2})|1\rangle +1/2|2\rangle +1/\sqrt{2}|3\rangle$ was postselected in $|f\rangle =1/\sqrt{2}|1\rangle -1/2|2\rangle +1/(2\sqrt{2})|3\rangle$ up to normalization, when the photon appeared at D1 (see the dashed vertical lines in figure \ref{fig:exp}).
In the postselection, $|f\rangle$, the plus (minus) sign for the coefficient of $|1\rangle$ ($|2\rangle$) was due to that the port of D1 was set to be the bright (dark) port of the interferometer composed of $|1\rangle$ and $|3\rangle$ ($|2\rangle$ and $|3\rangle$).
Note that this pre-postselection shows the same weak values as in equation (\ref{eq:wv_3box}).

According to the manner of weak measurement in \cite{WV_sym, WM1}, we experimentally observed the weak value of each path by HWP3, HWP4, and HWP5:
which path a photon had passed was written into it's polarization.
For example, we consider how to verify whether a photon has passed $|1\rangle$.
In this case, while the angles of the optical axes of HWP4 and HWP5 keep $0$ degree, the one of HWP3 is set so that the polarization is changed as $H$ $\rightarrow$ $V$, where $H$ ($V$) represents horizontal (vertical) polarization.
Note that an incident photon is horizontally polarized.
Then we can determine that the photon detected at D1 has (not) passed $|1\rangle$, if the polarization is $V$ ($H$).

\begin{figure}
 \begin{center}
	\includegraphics[scale=0.6]{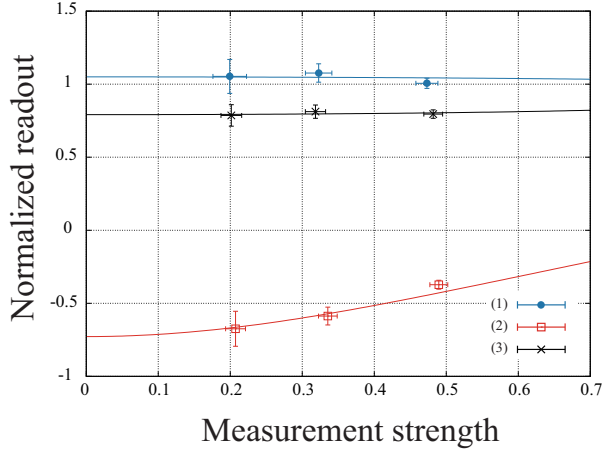}
 \end{center}
	\caption{The result of weak measurement of (1)$|1\rangle\langle 1|$, (2)$|2\rangle\langle 2|$, and (3)$|3\rangle\langle 3|$.
The value at $G=0$ corresponds to each weak value.
The error bars come from the statistical errors.
According to our fitting, the weak values are $\langle |1\rangle\langle 1|\rangle_{\bf w}=1.05$, $\langle |2\rangle\langle 2|\rangle_{\bf w}=-0.73$, and $\langle |3\rangle\langle 3|\rangle _{\bf w}=0.79$.
}
\label{fig:wm}
\end{figure}

Such a measurement completely disturbs the path state due to the strong correlation with the polarization state.
To achieve weak measurement, we should set the angle of HWP3 so that the polarization of a photon is almost $H$, namely, ${\rm cos}\alpha|H\rangle +{\rm sin}\alpha|V\rangle$ \ ($\alpha \sim 0$), by which the correlation between the path and the polarization becomes weak.
Using HWP6 and PBS3, we detect a photon in the polarization basis of $|\pm\rangle={\rm cos}(\alpha\pm\pi/4)|H\rangle+{\rm sin}(\alpha\pm\pi/4)|V\rangle$, and estimate their probability distribution, $P(+|f)$ and $P(-|f)$.
Then the normalized readout given by
\begin{eqnarray}
	R(+|f) = [P(+|f)-{\rm sin}^2\beta]/({\rm cos}^2\beta -{\rm sin}^2\beta)
\end{eqnarray}
corresponds to the pointer of measurement, and $G={\rm cos}^2\beta -{\rm sin}^2\beta$ is the measurement strength, where $\beta=\pi/4-\alpha$ \cite{WV_sym}.
Actually the normalized readout shows ${\rm Re}\langle|1\rangle\langle 1|\rangle _{\bf w}$ as $G\rightarrow 0$.
In this way we can also observe the other weak values, and figure \ref{fig:wm} shows our experimental result of weak measurement.
Their estimated weak values almost agree with the values in equation (\ref{eq:wv_3box}).

\begin{figure}
 \begin{center}
	\includegraphics[scale=0.55]{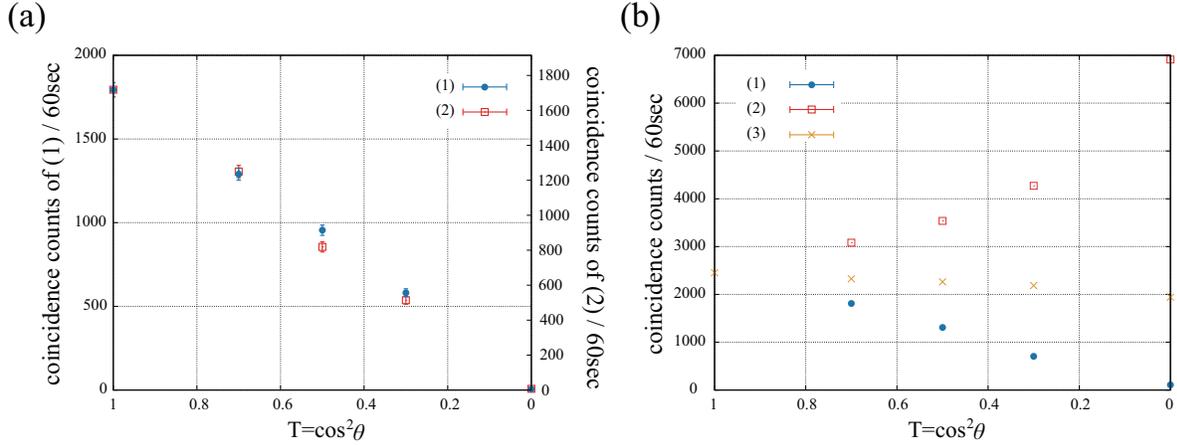}
 \end{center}
	\caption{(a)The decrease of detecting photons in relation to the expected transmission probability, $T$, for the individual path of (1)$|1\rangle$ and (2)$|2\rangle$.
The error bars represent the statistical errors.
(b)The coincidence counts when the loss with the transmission probability, $T$, was set in (1)$|1\rangle$ only, (2)$|2\rangle$ only, and (3)both $|1\rangle$ and $|2\rangle$.
Their counts at $T=0$ (i.e. no losses) are common.
The statistical errors are smaller that the plots.
}
\label{fig:result}
\end{figure}

We achieved the loss of photons by HBS and PBS represented by the dashed box in figure \ref{fig:exp}, which was prepared for the paths of $|1\rangle$ and $|2\rangle$.
Since an incident photon is horizontally polarized, the polarization can change into ${\rm cos}\theta|H\rangle +{\rm sin}\theta|V\rangle$ by using the HBS; due to the following PBS, they perform the loss with the transmission probability, $T={\rm cos}^2\theta$.
Note that the angles of the optical axes of HWP3, HWP4, HWP5, and HWP6 keep $0$ degree during the demonstration.
To verify whether our setup really achieves the photon loss, we observed how the detecting counts of photons were changed in response to $\theta$ for each path, blocking the other paths.
The results are plotted in figure \ref{fig:result} (a);
the decreases of the coincidence counts agree with the expected transmission probability within a few percents (i.e. they decreased linearly).

Then, in the pre-postselection of the path state, we observed the coincidence counts of photons by adjusting the HWP for the loss.
In figure \ref{fig:result} (b), the plots of (1) represents the case when the loss was set only in the path of $|1\rangle$.
The decrease agrees with the transmission probability, $T$, within a few percents: in agreement with $\langle|1\rangle\langle 1|\rangle _{\bf w}=1$, the detection probability decreases as if photons pass $|1\rangle$ with the probability 1.
The plots of (2) shows the case when the loss was placed only in $|2\rangle$.
Because the weak value of $|2\rangle\langle 2|$ was negative, the detection counts increased according as equation (\ref{eq:prob}).
Then we observed the coincidence counts when the same losses were set in both $|1\rangle$ and $|2\rangle$, whose results are plotted as (3) in figure \ref{fig:result} (b).
The decreased counts of (1) increased to the original counts of no losses (i.e. $T=1$) nearly: the loss at $|1\rangle$ was negated by the negative weak value of $|2\rangle\langle 2|$.
To be accurate, the counts of (3) seems to decrease gradually, although the counts should be constant irrespective of $T$ theoretically.
The cause of the slight decrease is that the loss at $|1\rangle$ was not completely negated by the one at $|2\rangle$.
Actually our estimated weak values are $\langle |1\rangle\langle 1|\rangle_{\bf w}=1.05$ and $\langle |2\rangle\langle 2|\rangle_{\bf w}=-0.73$ as shown in figure \ref{fig:wm}:
the absolute value of the negative value is slightly smaller than the positive weak value, and the negation was not completely achieved.
However, we were able to experimentally verify that the loss can be negated by a negative weak value.

\section{Negation provided by negative joint weak value}
\label{sec:joint}
Although we have discussed the case of a one particle, it can be applied to two or more particles, where joint weak values corresponding to joint probabilities are concerned:
the contribution of a unitary operator/particle loss involving multiple particles can also be estimated by using a joint weak value.
In this section, we would like to take up the specific case of two particles known as Hardy's paradox \cite{QPj1}.
Before getting into our subject, we review how a joint weak value has played an important role in the quantum paradox so far as in the previous section.

\begin{figure}
 \begin{center}
	\includegraphics[scale=0.9]{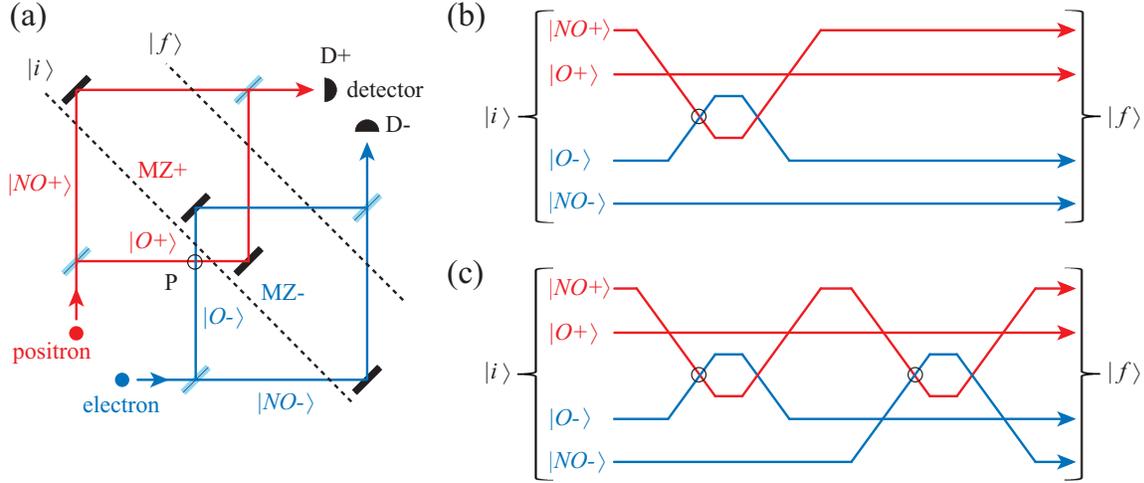}
 \end{center}
	\caption{(a)Hardy's paradox. The black circle, P, represents the overlapping of the paths such that a pair annihilation occurs when an electron and a positron simultaneously arrive at this point.
(b)In addition to the original overlapping point P, the paths of $NO+$ and $O-$ are overlapped between the pre-postselection to bring about a pair annihilation, which is represented by the black circle too.
(c)The overlapping between the pre-postselection is also added for the paths of $NO+$ and $NO-$.
}
\label{fig:jwv}
\end{figure}

In figure \ref{fig:jwv}, a positron and an electron are incident to their Mach-Zehnder interferometers, MZ+ and MZ-, respectively.
The paths of $|O+\rangle$ and $|O-\rangle$ overlap at P; if these particles simultaneously take the overlapping paths, a pair annihilation takes place.
Then the preselection can be represented by $|i\rangle = (|NO+, NO-\rangle+|NO+, O-\rangle+|O+, NO-\rangle)/\sqrt{3}$, as far as a pair annihilation does not occur.
The path lengths of each interferometer is adjusted so that the detection port becomes the dart port:
the detection on each interferometer means that the interferometer has been disturbed.
In this case, a coincidence detection corresponds to the postselection of the path state as $|f\rangle=(|NO+\rangle-|O+\rangle)(|NO-\rangle-|O-\rangle)/2$.
Actually we can easily find a finite probability of the coincidence detection.
Then, from the detection at $D+$ ($D-$), we can deduce that MZ+ (MZ-) has been disturbed by an electron (a positron) passing the overlapping path of $|O-\rangle$ ($|O+\rangle$).
If so, however, we were supposed to observe a pair annihilation and no coincidence count.
This is known as Hardy's paradox.

According to the pre-postselection, the joint weak values can be derived as follows \cite{QPj2, QPj3, QPj4},
\begin{eqnarray}
\langle |NO+, O-\rangle\langle NO+, O-|\rangle_{\bf w}=1, \ \langle |O+, NO-\rangle\langle O+, NO-|\rangle_{\bf w}=1, \nonumber \\
\langle |O+, O-\rangle\langle O+, O-|\rangle_{\bf w}=0, \ \langle |NO+, NO-\rangle\langle NO+, NO-|\rangle_{\bf w}=-1.   \label{eq:jwv}
\end{eqnarray}
These values represent the paradoxical argument well.
If we interpret $\langle |O+\rangle\langle O+|\rangle_{\bf w} = \langle |O+, O-\rangle\langle O+, O-|\rangle_{\bf w}+\langle |O+, NO-\rangle\langle O+, NO-|\rangle_{\bf w}=1$ as a probability, this value shows that a positron passes $|O+\rangle$ with certainty; we can drive the same result, $\langle |O-\rangle\langle O-|\rangle_{\bf w}=1$, about an electron too.
On the other hand, $\langle |O+, O-\rangle\langle O+, O-|\rangle_{\bf w}=0$ represents that there are no pair annihilations well.
Then, as shown in previous case of the quantum box problem, the negative weak value of $-1$ appears and satisfies the summation rule as follows,
\begin{eqnarray}
& &\langle |NO+, O-\rangle\langle NO+, O-|\rangle_{\bf w}+\langle |O+, NO-\rangle\langle O+, NO-|\rangle_{\bf w} \nonumber \\
& &+\langle |O+, O-\rangle\langle O+, O-|\rangle_{\bf w}+\langle |NO+, NO-\rangle\langle NO+, NO-|\rangle_{\bf w}=1.
\end{eqnarray}

We come back to the subject of the loss given by weak values.
As in section \ref{sec:intro}, when the joint weak value of a projector is 1, the quantum system behaves as in the corresponding eigenstate with probability 1.
According to the joint weak values in equation (\ref{eq:jwv}), for example, $\langle |NO+, O-\rangle\langle NO+, O-|\rangle_{\bf w}=1$ means that the joint probability for a positron and an electron to take the paths of $|NO+\rangle$ and $|O-\rangle$ respectively is $1$.
It can be verified by overlapping their paths between the preselection and the postselection so that it brings about a pair annihilation, when both of the particles take the paths simultaneously;
we show the schematic diagram in figure \ref{fig:jwv} (b).
In the context of the previous section, it corresponds to setting the loss with the transmission probability $T=0$ for $|NO+, O-\rangle$ (i.e. the shutter for two particles).
Then we can easily find that a pair annihilation takes place at the added overlapping point with certainty: the success probability of the postselection becomes 0.

If we also place an overlapping point for a pair annihilation between the paths of $|NO+\rangle$ and $|NO-\rangle$ as shown in figure \ref{fig:jwv} (c), the pair annihilation at the previously added overlapping point is negated (i.e. the probability of the postselection is recovered).
This is because of $\langle |NO+, NO-\rangle\langle NO+, NO-|\rangle_{\bf w}=-1$.
Actually joint weak values satisfy the equations (\ref{eq:uni}) and (\ref{eq:prob}), when there are unitary operations/losses whose effect involve multiple particles.
After all, the negative joint weak value of $-1$ well represents the negation against the positive joint weak value of $1$, equivalently, the joint probability of $1$.

\section{Conclusion}
\label{sec:con}
By postselection, an operation to a quantum system can be arbitrarily controlled, including the negation of the operation.
Although the control of an operation is accompanied by the cost of postselection, namely, discarding some samples failed to be postselected, we can assure that the expected operation has been performed whenever the postselection is achieved.
Such an operation originates from an interference forced by postselection;
in particular, the negation of an operation comes from a completely destructive interference.

Instead of following a time evolution of a quantum system, we have discussed the weak value, which gives us a simple manner to estimate the contribution of an operation between pre-postselection above mentioned.
As a result, we showed that the positive weak value of $1$ is equivalent to the probability $1$, as an operation is performed with certainty.
On the other hand, the negative weak value of $-1$ accomplishes the negation of the operation, since the weak value associated with the operation is to be $0$, namely, the probability $0$.

An operation we have treated includes an irreversible process; we have considered the loss of particles.
Actually we have experimentally verified how the loss of photons given by the weak value of $1$ was negated by the negative weak value of $-1$.
It is also notable that, in our case, the appearance of the weak values never need a certain `weak' condition as in weak measurement.
We hope that our demonstration contributes to not only the fundamental understanding of weak value but also the application for quantum information like a quantum circuit design between pre-postselection.

\section*{Acknowledgment}
On performing our experiment we thank R.~Ikuta and T.~Yamamoto for their cooperations.
K.~Y. also thanks L.~Vaidman for his helpful comments.
This work was supported by Core Research for Evolutional Science and Technology, Japan Science and Technology Agency (CREST, JST) JPMJCR1671 and Japan Society for the Promotion of Science (JSPS) Grant-in-Aid for Scientific Research(A) JP16H02214.

\section*{Appendix}
In our previous work \cite{WV_sym}, a half wave plate was placed on one of paths in figure \ref{fig:intro2} (a), and we considered how the linear-polarization was changed in response to the weak value of the projector of the path, especially the positive weak value of $1$ and the negative weak value of $-1$.
Then we found that the shift of the polarization given by the positive and negative weak values were symmetrical when the correlation between the polarization and the path state was weak enough as follows.

In figure \ref{fig:intro2} (a), we assume that the polarization of an incident photon is initially horizontal, $|H\rangle = (1,0)^{t}$, and a unitary operation is placed in only one of the paths.
We also suppose that the unitary operation gives the rotation, $\hat{U}(\phi)$, as follows,
\begin{eqnarray}
  \hat{U}(\phi) = \left(
    \begin{array}{ccc}
      {\rm cos}\phi & {\rm sin}\phi \\
      -{\rm sin}\phi & {\rm cos}\phi
    \end{array}
  \right)
  \sim \left(
    \begin{array}{ccc}
      1 & \phi \\
      -\phi & 1
    \end{array}
  \right) \ \ (\phi\sim0).
\end{eqnarray}
The approximation corresponds to when the correlation between the polarization and the path is weak, since the polarization is almost horizontal no matter which path the photon has passed.

According to equation (\ref{eq:uni}), when the weak value is $1$, this rotation is directly given: the direction of the polarization shifts to $\phi$.
On the other hand, the operation when the value is $-1$ is given by
\begin{eqnarray}
2-\hat{U}(\phi) = \left(
    \begin{array}{ccc}
      2-{\rm cos}\phi & -{\rm sin}\phi \\
      {\rm sin}\phi & 2-{\rm cos}\phi
    \end{array}
  \right)
  \sim \left(
    \begin{array}{ccc}
      1 & -\phi \\
      \phi & 1
    \end{array}
  \right) 
\sim \hat{U}(-\phi) \ (\phi\sim0),
\end{eqnarray}
which is symmetrical to that one given by the weak value of $1$.
Although these symmetrical shifts are satisfied under the condition of the weak correlation between the polarization and the path, the negation provided by the negative weak value of $-1$ against the positive weak value of $1$ is approved without such a condition as in figure \ref{fig:intro} (c).

\end{document}